\def\td{\mbox{d}}
\def\d{\partial}

\def\<{\langle}
\def\>{\rangle}

\documentclass[aps,amsmath,amssymb,floatfix,showpacs]{revtex4}

\usepackage{epsf}
\begin{document}

\title{Detailed balance has a counterpart in non-equilibrium steady states}
\author{R. M. L. Evans}
\affiliation{University of Leeds, LS2 9JT, U.K.}
\date{11 October 2004}
\begin{abstract}

When modelling driven steady states of matter, it is common practice either to
choose transition rates arbitrarily, or to assume that the principle of 
detailed balance remains valid away from equilibrium. Neither of those 
practices is theoretically well founded. Hypothesising ergodicity 
constrains the transition rates in driven steady states to respect relations 
analogous to, but different from the equilibrium principle of detailed balance. 
The constraints arise from demanding that the design of any model system contains 
no information extraneous to the microscopic laws of motion and the macroscopic 
observables. This prevents over-description of the
non-equilibrium reservoir, and implies that not all stochastic equations of 
motion are equally valid. The resulting recipe for transition rates has many 
features in common with equilibrium statistical mechanics.

\end{abstract}


\maketitle

\section{Introduction}

I address the question, what is the appropriate stochastic equation of
motion to use when modelling a driven steady state (including chaotic and
fluctuating steady states) such as that of a fluid under continuous shear
flow? At equilibrium, the solution is well understood. To generate
configurations consistent with the equilibrium ensemble, one may use any
equation of motion that respects the principle of detailed balance, which
is a constraint on ratios of forward and reverse transition rates. That
condition ensures that every thermally-driven flux is balanced by an equal
and opposite flux. For non-equilibrium systems in continuously driven
steady states, no such guidance is hitherto available in choosing an
equation of motion consistent with the {\em mechanically} (externally)
driven fluxes, so arbitrary choices are often made. The aim of this work 
is to eliminate arbitrariness, and determine what transition rates are implied 
by the macroscopic state of the non-equilibrium system, i.e. its mean energy 
and flux, combined with our knowledge of the microscopic laws of physics. The 
objective is to use {\em only} the information that is available, without 
unwittingly introducing any arbitrariness, deriving from personal prejudices. 
The method for keeping the amount of information constant throughout the 
calculation is Jaynes' information-theoretic method of maximum 
entropy inference \cite{Jaynes57,Shannon}(MaxEnt), which is often misunderstood
in the context of non-equilibrium thermodynamics, despite recent notable 
achievements \cite{Dewar03}. It has been successfully used to derive fluctuation 
theorems \cite{Maes99} and linear transport theory \cite{JaynesBook}, and to 
explain self-organised criticality \cite{Dewar03}. 

Jaynes gives a nice explanation of maximum entropy inference in his original 
paper on the subject \cite{Jaynes57}, where he uses the method to 
re-derive equilibrium statistical mechanics without the need for many 
microscopic details that had previously been considered necessary. The 
application of the method to equilibrium systems is uncontroversial. However, 
the history of non-equilibrium information theory can be confusing because 
it has been used in so many different ways, some of them exact, some only
approximate. In fact, information theory itself is not a physical theory, but 
a mathematical method, providing a logical structure. Some physical input is 
required if such a method is to make physical predictions. If one throws 
away too much relevant information about some non-equilibrium system before 
applying MaxEnt, it will still provide answers, but they will be inaccurate. 
For instance, using the method to minimise the information content of the 
momentum distribution in a non-equilibrium gas, although efficacious, is not an
exact method, as was recently shown \cite{Hyeon}. In fact, there is no 
justification for discarding all information content except for some averaged 
features. Indeed, particles possess their individual velocities for a reason: 
they have each come from somewhere, and are going somewhere, and their journeys 
will affect the trajectories of other particles. These facts are relevant to the 
physics of a non-equilibrium system, and lead to temporal correlations. 

At the other extreme, if one retains all the details of a system's phase-space 
trajectory, allowing no stochastic input (e.g.~from a 
reservoir), then MaxEnt becomes a null procedure, since it is asked to choose 
the most likely distribution from a choice of only one physical scenario - a 
delta function distribution of trajectories. Such a null procedure may be 
regarded as an extreme case where 
MaxEnt can correctly ``predict" any and all physics. There is thus no reason in 
principle why MaxEnt should be expected to fail in non-equilibrium situations, 
if we ask it the right questions.

The choice of the prior set of options that is presented to MaxEnt is of crucial 
importance. It should be a set of physical paths through phase space, that each 
obeys Newton's laws, so that all physics (the Navier-Stokes equation, long-range 
correlations, etc.) is respected {\it a priori}. MaxEnt then tells us which of 
these trajectories is 
most likely to be chosen, under the influence of a non-equilibrium reservoir 
that is coupled to the system but uncorrelated with it\footnote{The ``reservoir" 
may in fact stand for the rest of the ensemble of systems, as in the equilibrium 
derivation of the Gibbs ensemble.}. This is the application of information 
theory that should be understood here and in Refs.~\cite{PRL,sardinia}.
I derive its implications for transition rates. 

Alternatively, as is often done in theoretical modelling, one can settle for a 
physically imprecise prior set of dynamical rules --- such as Brownian 
particles, or a discrete state space, or discrete time steps --- so that things 
become easy to solve. Then applying the methods below will, accordingly, yield 
only approximate physics, but at least one will know exactly what information 
went into the simplified model. Such an application of the present theory would 
yield transition rates that are somewhat arbitrary, due to the arbitrariness of 
the prior rates that are chosen. However, it will provide strictly {\em the 
least} arbitrary model. Such a model will be derived in section 
\ref{applications} for a stochastically hopping particle, that demonstrates some 
features of the method. It is important to realise that the approximations 
introduced in that section are only for expediency in that particular model. The 
general derivation of the method for obtaining transition rates from prior 
dynamical rules combined with non-equilibrium macroscopic observables, presented 
in sections \ref{presentation} and \ref{variants}, remains exact.

The conditions derived here for macroscopically driven steady states are
analogous to the equilibrium principle of detailed balance. Like
detailed balance, the conditions are not sufficient to 
completely determine the microscopic transition rates, but are necessary to be 
satisfied by any equation of motion that generates an unbiased ergodic driven 
steady-state ensemble. The derivation of detailed balance relies on two 
assumptions: time-reversal symmetry of the microscopic laws of motion, and the
ergodic hypothesis which implies that a heat reservoir can be characterised by 
the Boltzmann distribution with temperature as the only parameter. Similarly, the
non-equilibrium conditions assume the same microscopic laws that govern 
equilibrium motions (therefore implicitly requiring microscopic time-reversal 
symmetry, broken only by imposition of the macroscopic flux), and 
rely also on a hypothesis of ergodicity implying that the driven reservoir is 
fully characterised by its macroscopic observables (mean energy and flux).
Many quiescent systems (those without fluxes) are at thermodynamic
equilibrium, but exceptions include glasses \cite{glasses}, granular media
\cite{Wittmer,Fragile}, and certain cellular automata \cite{cellular}, in which
the ergodic hypothesis and/or microscopic reversibility fails.
Boltzmann's law and the principle of detailed balance apply only to that
class of quiescent systems that are, by definition, at equilibrium. That 
class of systems has of course proved to be large, significant and interesting.
Similarly, not every non-equilibrium steady state should be expected to respect 
the conditions presented here; exceptions include traffic flow and fluids of 
molecular motors, in which the constituents violate time reversal symmetry. The
ergodic hypothesis may also fail in some systems, implying that hidden 
information that is not apparent in the macroscopic observables is nonetheless
significant. However, it is anticipated that the ergodicity criteria are respected
by the transition rates of many macroscopically driven systems, defining a
special and important class.

The method outlined in section \ref{presentation} was presented in a
recent Letter \cite{PRL}. It is explained here in more detail, and the
analysis extended to an alternative non-equilibrium ensemble in section
\ref{variants}. The method is demonstrated in section \ref{model} where rates 
are derived for the stochastic transitions of a particle hopping in a 
non-trivial energy landscape, subject to a driving force. Applications to 
other models are also discussed in section \ref{applications}.

\section{The method}
\label{presentation}

\subsection{Information entropy}

Using Jaynes' interpretation of Gibbs' entropy \cite{Jaynes57}, it is
possible to make a ``Maximum Entropy Inference" \cite{Jaynes57,Dewar03}
to assess the probability that a system, subject to random 
influences, (whether at equilibrium or not) takes a particular trajectory 
$\Gamma_0$ through its phase space, thus allowing us to assess the {\em 
reproducible part} \cite{Jaynes57,JaynesBook} of the system's motion. The 
recipe for the probability $p(\Gamma_0)$ of trajectory $\Gamma_0$ is to 
maximize the Shannon entropy, or information entropy,
\begin{equation}
\label{information}
	S_I \equiv -\sum_{\Gamma} p(\Gamma) \ln p(\Gamma)
\end{equation}
subject to constraints that some averaged properties of the trajectories
conform with our knowledge of the macroscopic features such as
mean energy, volume, flux etc.

In principle, this formalism gives us a full solution of the statistics of
any ensemble, be it at equilibrium or not. In the absence of any
macroscopic fluxes (i.e.~at equilibrium), the prescription reduces to a
maximization of the Gibbs entropy with respect to a distribution of
instantaneous states rather than trajectories, yielding Boltzmann's law.
In the non-equilibrium case, MaxEnt gives us the probability of 
an entire trajectory $\Gamma_0$. It would be more useful to have a formula 
for the probability of a short segment of the trajectory, a single transition
from a state $a$ to a subsequent state $b$. Such a transition probability
is what we require for designing a stochastic model or simulation. This
would allow us to generate trajectories belonging to the 
non-equilibrium ensemble. Let us now derive that formula. We begin with
some trivial calculations to establish notation.

\subsection{Prior probability}
\label{pp}

At any instant $t$, the entire state of a system is represented
classically by its phase-space position vector $x(t)$. This is a
high-dimensional vector specifying the positions and momenta of all the
particles constituting the system. As time progresses from the beginning
$t=0$ to the end $t=\tau_0$ of the experiment or simulation, $x(t)$ traces
out a trajectory $\Gamma_0$ through phase space. It will prove useful to
label each probability distribution function with a subscript indicating
the duration of the trajectories to which it applies thus:
$p_{\tau_0}(\Gamma_0)$. For a deterministic system with definite initial
conditions, only one trajectory is possible, so the probability
distribution is a delta function. In the presence of randomness, such as
coupling to a reservoir of systems with similar properties, the
distribution is finite for all trajectories that respect some prior
dynamical rules, such as conservation of momentum for all internal degrees
of freedom not directly coupled to the reservoir.

In the {\em absence} of any posterior constraints other than normalization, 
\begin{equation}
\label{norm}
  \sum_{\Gamma} p_{\tau_0}(\Gamma) = 1,
\end{equation}
all trajectories of a given duration $\tau_0$ have equal {\it a priori}
probability. That is not an independent postulate, but is embodied in the
maximum entropy principle of information theory
\cite{Jaynes57,JaynesBook}, since the entropy-maximizing distribution is
given by
\begin{equation}
\label{prior}
  \frac{\d}{\d p_{\tau_0}(\Gamma_0)} \sum_{\Gamma}
  \Big\{ -p_{\tau_0}(\Gamma) \ln p_{\tau_0}(\Gamma) + \lambda\,
  p_{\tau_0}(\Gamma) \Big\} = 0
\end{equation}
with a Lagrange multiplier $\lambda$ chosen for consistency with
Eq.~(\ref{norm}). Equation~(\ref{prior}) is solved by
$p_{\tau_0}(\Gamma_0)=\mbox{constant}$, indicating that the unconstrained
(`prior') set of trajectories of a given duration have equal probability.

\subsection{Equilibrium ensemble}

We now impose a posterior constraint, and calculate the statistical
properties of that sub-set of trajectories that respect the constraint.
Let us not necessarily conserve the energy $E$ of the system at each
instant (since we allow energy exchange with a reservoir), but rather demand 
that its time-average over the whole trajectory
$\Gamma_0$ is fixed at $E_0$. We shall use a bar to indicate
time averages, so that
\begin{equation}
\label{meanE}
  \overline{E_{\Gamma_0}} \equiv \frac{1}{\tau_0} \int_0^{\tau_0} 
  E_{\Gamma_0}(t)\: \td t = E_0.
\end{equation}

Let us divide the trajectory $\Gamma_0$ into shorter segments $\Gamma$, each 
of duration $\tau$. Then the constraint on the time-averaged energy may be 
written
\begin{equation}
\label{canonical}
  \tau \sum_\Gamma \overline{E_\Gamma} = \tau_0 E_0.
\end{equation}
Assuming ergodicity, time-averages are equivalent to ensemble-averages in
the limit $\tau_0/\tau\to\infty$. So this constraint, for a time-average
on $\Gamma_0$, defines the equilibrium canonical ensemble for $\Gamma$. In
other words, the conditional probability 
\mbox{$p_{\tau}(\Gamma |\overline{E_{\Gamma_0}}=E_0)$} of 
encountering a particular trajectory segment
$\Gamma$ of duration $\tau$, given that the whole trajectory has a
time-averaged energy $E_0$, is found by maximizing the
information entropy for $\Gamma$ subject to Eq.~(\ref{canonical}). The
maximization involves Lagrange multipliers $\beta$ for this energy
constraint, and $Z^{-1}$ for the normalization constraint, and yields
\[
  p_{\tau}(\Gamma | E_0 ) = Z^{-1} \exp(-\beta \overline{E_\Gamma})
\]
where the condition $\overline{E_{\Gamma_0}}=E_0$ is represented for brevity 
by $E_0$. As expected, this is Boltzmann's law, and we interpret the Lagrange
multipliers as the temperature parameter $\beta=1/k_BT$ and partition
function \mbox{$Z=\sum_\Gamma \exp(-\beta \overline{E_\Gamma})$}.

\subsection{Transition rates}

A transition rate, for any transition between states $a\to b$ say, can be
written as a conditional probability. If we consider a trajectory segment
$\Gamma'$ of duration $\Delta t$, representing the
transition $a\to b$, then the transition rate at some time, which we may
define without loss of generality to be $t=0$, is
\begin{equation}
\label{priorrate}
  \omega^{\rm prior}_{\Gamma'}=p_{\Delta t}(\Gamma' | x(0)=a )/\Delta t. 
\end{equation}
This is the probability (per unit time) of encountering the trajectory
\mbox{$\Gamma'\equiv\{x(0)=a, x(\Delta t)=b\}$}, {\em given} that we begin
at $a$. Equation~(\ref{priorrate}) gives the {\em prior} rate of the
particular transition. The rate in the equilibrium ensemble is given by a
probability subject to {\em two} conditions:
\begin{equation}
\label{eqrate}
  \omega^{\rm eq}_{\Gamma'} 
  = p_{\Delta t}(\Gamma' | a, E_0) /\Delta t
\end{equation}
where the condition $x(0)=a$ is represented for brevity by $a$. To re-cap,
Eq.~(\ref{eqrate}) defines the probability of encountering trajectory
segment $\Gamma'$ (a transition $a\to b$) given that we are in state $a$,
and that the entire trajectory $\Gamma_0$ of duration $\tau_0$ will turn out 
to have a mean energy $E_0$.

\subsection{Driven ensemble}
\label{driven}

We have looked so far at the prior phase-space trajectories, and those for
systems in the equilibrium ensemble. Our goal is to determine the
transition rates appropriate to a non-equilibrium ensemble, for which
there is an imposed flux $J$. Again, we should not over-constrain the
dynamics. Let us allow the flux to fluctuate, and demand only that the
dynamics will result in some finite value $J_0$ of the flux time-averaged over
the whole trajectory:
\begin{equation}
\label{meanj}
  \overline{J_{\Gamma_0}} \equiv \frac{1}{\tau_0} \int_0^{\tau_0} 
  J_{\Gamma_0}(t)\: \td t = J_0.
\end{equation}
We ask, what is the probability, in time $\Delta t$, of encountering the
transition $\Gamma'=\{a\to b\}$, given that we begin in state $a$, and
that the dynamics will eventually conspire to produce a mean flux
$J_0$ and energy $E_0$? Again we relate this
conditional probability to a transition rate:
\begin{equation}
\label{drivenrate}
  \omega^{\rm dr}_{\Gamma'} = p_{\Delta t}(\Gamma' | a, J_0, E_0)/\Delta t.
\end{equation}

\begin{figure}
%% \displaywidth\columnwidth
  \epsfysize=5cm
  \begin{center}
  \leavevmode\epsffile{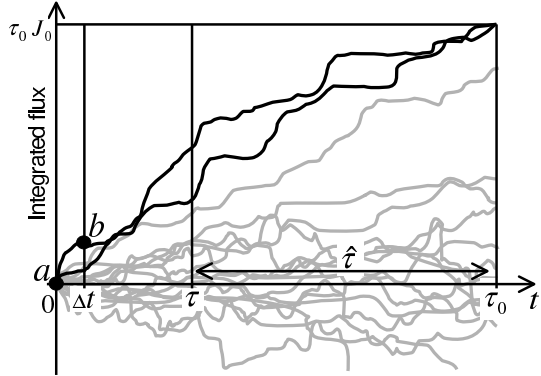}
  \caption{A sample of phase-space trajectories of duration $\tau_0$. One 
phase-space coordinate (cumulative flux) is shown as a function of time 
$t$. At equilibrium, most trajectories lie near the horizontal axis. The
sub-set of trajectories shown in black belongs to the microcanonical-flux 
ensemble. Their time-averaged flux is $J_0$. The probability 
of a transition, in microscopic time $\Delta t$, between the phase-space 
points $a$ and $b$ shown, is enhanced for this sub-set, relative to its 
equilibrium likelihood.}
\label{trajectories}
\end{center}
\end{figure}

Fig.~\ref{trajectories} depicts some of the trajectories that have been
discussed. Time $t$ is shown on the horizontal axis, and all trajectories
have a total duration $\tau_0$. The vertical axis represents the
phase-space coordinates though, of course, this is a reduced
representation of the vastly high-dimensional phase space, since it has
been projected onto a single axis. For definiteness, let us say that this
axis represents integrated flux, i.e. the flux that the system has accrued
since $t=0$. We must imagine that all the other coordinates required to
fully describe the state of the system, are on axes perpendicular to the
page.

A sample of trajectories representing the equilibrium distribution is
shown (in grey and black). These trajectories are concentrated close to
the time axis (zero flux). If another axis measuring instantaneous energy
$E(t)$ were constructed perpendicular to the page, then the density of
trajectories would be exponentially distributed along that axis, by
Boltzmann's law. Equation (\ref{eqrate}) gives the frequency of observing
a particular trajectory segment shown in Fig.~\ref{trajectories} (the
single transition $a\to b$) of microscopic duration $\Delta t$, given that
we are currently (at $t=0$) in state $a$, and that the whole trajectory
belongs to this equilibrium set. Equation (\ref{drivenrate}) asks for the
frequency with which that trajectory segment $\{a\to b\}$ occurs in the
sub-set of trajectories shown in black in Fig.~\ref{trajectories}, for
which a given integrated flux will be accumulated by time $t=\tau_0$. This
sub-set of trajectories is the driven ensemble.

Note that we shall not require $\Delta t$ to vanish. The discussion will 
cover discrete-time processes for which the microscopic time step is 
$\Delta t\equiv1$, as well as continuous-time dynamics for which 
$\Delta t\to0$.

\subsection{Bayesian evaluation}

To mathematically manipulate conditional probabilities, we appeal to
Bayes' theorem. It states that the joint probability of two outcomes $X$
and $Y$ both occurring, given a third fact $Z$, may be written in two
equivalent ways:
\begin{equation}
\label{Bayes}
  p(X|Z) p(Y|X,Z) = p(Y|Z) p(X|Y,Z)
\end{equation}
where $p$ is simply used to mean `probability' for any event
(appropriately normalised), as opposed to a particular distribution
function. We can now assign the following meanings: $X$ is the fact that
the transition $a\to b$ takes place within $\Delta t$, represented by the
trajectory $\Gamma'$; $Y$ says that the flux has a mean value
$\overline{J_{\Gamma_0}}=J_0$ averaged over the entire duration $\tau_0$; and 
$Z$ is
the combined statement that the initial state at $t=0$ is $a$ and that the
trajectory's time-averaged energy will be $\overline{E_{\Gamma_0}}=E_0$. Thus,
Eq.~(\ref{Bayes}) expresses the probability of the transition taking place
within $\Delta t$ {\em and} the flux averaged over $\tau_0$ being
$J_0$, for the given initial state and average energy. It is
re-written thus:
\begin{equation}
  p_{\tau_0}(J_0|a,E_0)\, p_{\Delta t}(\Gamma'|a,J_0,E_0) =
  p_{\Delta t}(\Gamma'|a,E_0)\, p_{\tau_0}(J_0|a,\Gamma',E_0).
\end{equation}
Notice that it is redundant to specify the two conditions $(a,\Gamma')$,
since the trajectory segment $\Gamma'$ is the transition $a\to b$ which
includes the initial state $a$. Substituting from Eqs.~(\ref{eqrate}) and
(\ref{drivenrate}) yields a theorem for transition rates in the driven
steady-state ensemble,
\begin{equation}
\label{relation}
  \omega^{\rm dr}_{a\to b} = \omega^{\rm eq}_{a\to b} \, 
  \lim_{\tau_0\to\infty} \frac{p_{\tau_0}^{\rm eq}(J_0 | a\to b)}
  {p_{\tau_0}^{\rm eq}(J_0 | a)}.
\end{equation}
Notice that all quantities on the RHS of Eq.~(\ref{relation}) are defined
{\em at equilibrium}, not on the driven ensemble. This is indicated by the
superscript `eq', which is equivalent to the condition fixing
$\overline{E_{\Gamma_0}}$, the time-averaged energy. Equation~(\ref{relation})
tells us that the transition rate in the driven ensemble is given by the
transition rate in the equilibrium ensemble, multiplied by an enhancement
or attenuation factor. We shall see below that the theorem makes intuitive
sense.
 
Given that the dynamics must be consistent with the macroscopically
observable mean energy and flux, and with the same microscopic laws of 
motion that hold sway in an equilibrium system, MaxEnt yields an unbiased 
description ofthe dynamics, and thereby constrains the system the least. 
Equation(\ref{relation}) specifies explicitly the dynamical rules implied
byMaxEnt. How do we know that Eq.~(\ref{relation}) constrains the dynamics
the least? It does, because all the quantities on the RHS are defined for
the maximum-entropy ensemble at equilibrium, i.e.~without the extra
constraint on the flux. Given that we start with an unbiased set (the
equilibrium ensemble), Bayes theorem gives us the least biased set subject
to the extra posterior constraint.

Let us examine the enhancement factor in Eq.~(\ref{relation}) in detail.
It is a ratio of conditional probabilities for encountering a flux
$J_0$ in the equilibrium ensemble. Of course, we do not expect
a system at equilibrium to exhibit any net flux, averaged over its whole
trajectory. The chance of such a flux arising spontaneously at equilibrium
is vanishingly small (as $\tau_0\to\infty$). So the RHS of
Eq.~(\ref{relation}) is the ratio of two vanishingly small terms. However
unlikely it may be for an equilibrium system to spontaneously exhibit the
desired macroscopic flux, we ask, how much would that probability be
enhanced as a result of the putative transition $a\to b$? If the dynamics
of the transition itself contributes some flux to the trajectory, it is favoured 
by the enhancement factor. The factor also favours transitions to configurations
that give a greater than average probability of subsequently obtaining the 
desired flux, for the given starting point. If the new state $b$ is more likely 
to initiate high-flux trajectories, then the transition rate is boosted over and 
above the equilibrium rate.

We shall examine the implications of Eq.~(\ref{relation}) in some examples,
but firstly let us interpret the meaning of its derivation.
Imagine that a lazy physicist wishes to collect data from a driven steady
state, such as continuous shear flow of a complex fluid. Our physicist has
a computer program that simulates the fluid at equilibrium (with free or
frictionless boundaries, say), and is too lazy to write a new program that
simulates shear.
Instead, (s)he runs the equilibrium simulation, in the hope that it will
spontaneously exhibit shear flow. It does not. So the dilettante updates
the program's random number generator and runs it again. Having tenure and
little imagination, the physicist repeats this process countless times
until, one day, the fluid fluctuates into a state of sustained shear flow.
The delighted simulator records this fluke, but continues the project for
many more years until a large number of such accidents have been observed,
exhibiting the same shear rate. Finally, the researcher discards an
enormous set of simulated trajectories, and publishes only that subset
which happened to perform the desired shear. On analysing this subset of
trajectories, one might expect to observe the equilibrium transition rates
that were coded into the algorithm. But this is a biased data set, subject
to an {\it a posteriori} constraint of shear flux $J_0$. So
this sub-set of the equilibrium ensemble exhibits exactly the transition
rates specified by Eq.~(\ref{relation}). Although the programmer has
published a biased account of the equilibrium simulations, there was no
unwarranted or subjective bias other than the flux constraint, hence the
project was a success in producing the physics of shear flow. 

Note that,
despite extracting a sub-ensemble from the equilibrium ensemble, the lazy
physicist has {\em not} produced a near-equilibrium approximation. The
``sub-ensemble dynamics" of Eq.~(\ref{relation}) has features
qualitatively different from the equilibrium dynamics.

In section \ref{applications}, I shall use some examples to demonstrate 
the correctness of the physics generated by sub-ensemble dynamics 
(Eq.~(\ref{relation})). Before doing so, in section \ref{variants}, I 
develop a useful variant of Eq.~(\ref{relation}), analogous to an 
alternative thermodynamic ensemble.

\section{Alternative dynamic ensembles}
\label{variants}

\subsection{Microcanonical-flux ensemble}
\label{micro}

Equation (\ref{relation}) gives the frequency of observing a particular
trajectory segment (e.g.~a single transition $a\to b$) of microscopic
duration $\Delta t$, in the driven ensemble which is a sub-set ofall
trajectories, shown in black in Fig.~\ref{trajectories}. These
trajectories lie in the extreme tails of the equilibrium distribution.
Note that they have common end points, since we have specified the exact
net flux that must flow during the duration of the experiment. Any nearby
trajectories, that do not have {\em exactly} the specified flux, do not
contribute to the quantities appearing in Eq.~(\ref{relation}). Even very
nearby trajectories are completely discarded by Eq.~(\ref{relation}). This
can be seen by re-writing the probability of thespecified flux 
$J_0$ as a sum over trajectories $\Gamma_0$ with
fluxes $\overline{J_{\Gamma_0}}$, so that Eq.~(\ref{relation}) becomes
\begin{eqnarray}
\label{dynmicro}
  \frac{\omega^{\rm dr}_{a\to b}}{\omega^{\rm eq}_{a\to b}} &=& \frac{
  \int\td J\: p_{\tau_0}^{\rm eq}(J|a\to b)\:\delta(J-J_0)}
  {\int\td J\: p_{\tau_0}^{\rm eq}(J|a)\:\delta(J-J_0)}	\\
  &=& \frac{\sum_{\Gamma_0} p_{\tau_0}^{\rm eq}(\Gamma_0|a\to b)\, 
  \delta(\overline{J_{\Gamma_0}}-J_0)}
  {\sum_{\Gamma_0}p_{\tau_0}^{\rm eq}(\Gamma_0|a)\, 
  \delta(\overline{J_{\Gamma_0}}-J_0)}.	\nonumber
\end{eqnarray}
Here, the Dirac delta functions kill all trajectories with anything but
the exact net flux $J_0$. This can be a disadvantage for
practical applications of the formula. (The lazy physicist, discussed above, 
must discard data even from simulations that produce {\em nearly} the right 
flux.) An alternative expression is now
derived, that samples trajectories with less stringent conditions on their
eventual flux content.

\subsection{Canonical-flux ensemble}
\label{canon}

In equilibrium statistical mechanics, the constraint of energy conservation 
is relaxed by dividing the isolated microcanonical system into a 
relatively small sub-section, defining the canonical system, and the large 
remainder, known as the 
reservoir. Similarly, we shall relax the strict constraint on the 
time-averaged flux, by dividing the total trajectory of duration $\tau_0$ 
into a part (see Fig.~\ref{trajectories}) of duration $\tau$ (where 
$\Delta t\ll\tau\ll\tau_0$), whose 
properties are examined in detail, and the large remaining part of 
duration $\hat{\tau}\equiv\tau_0-\tau$, for which the system's motion is
uncorrelated with the earlier trajectory segment.

We may express the conditional probability $p_{\tau_0}(J_0|a)$
of a net flux $J_0$ in the full duration $\tau_0$, as an
integral over all possible fluxes during the interval $\tau$ thus:
\begin{equation}
\label{workings1}
  p_{\tau_0}^{\rm eq}(J_0|a) = 
  \int_{-\infty}^\infty \td J\: p_{\tau}^{\rm eq}(J|a)\,
  p_{\hat{\tau}}^{\rm eq}(\hat{J}|a, J)
\end{equation}
where $p_{\hat{\tau}}(\hat{J}|a, J)$ is the probability of an appropriate
flux $\hat{J}$ during interval $\hat{\tau}$ given that the system began in
state $a$ at $t=0$, and then flowed with mean flux $J$ for the duration
$\tau$. The required flux $\hat{J}$ is given by
\begin{equation}
\label{fluxsum}
  \tau J + \hat{\tau}\hat{J} = \tau_0 J_0.
\end{equation}
Given that $\tau$ exceeds any correlation time $\tau_{\rm cor}$, the
probability $p_{\hat{\tau}}(\hat{J}|a,J)$ becomes independent of $a$,
because the system has forgotten its initial state by the time $\tau$ at
which the interval $\hat{\tau}$ commences. In fact, at time $\tau$, the
system is in a state drawn at random from the driven steady-state
ensemble, since the integral in Eq.~(\ref{workings1}) is dominated by
$J\approx J_0$. So we may make the replacement
\begin{equation}
\label{replacement}
  p_{\hat{\tau}}(\hat{J}|a, J) \to  
  \sum_c p^{\rm dr}(c) \:p_{\hat{\tau}}^{\rm eq}(\hat{J}|c)
\end{equation}
where $p^{\rm dr}(c)$ is the steady-state distribution of instantaneous 
microstates in the driven ensemble. Not only is the above
formula independent of the initial state $a$, it actually takes a universal
(exponential) form as a function of $J$, as shown in 
Appendix \ref{canonapp} using the
theory of large deviations. This is because the extremely unlikely value of
the flux, $\hat{J}$, is the result of many unlikely realisations of the flux
during the many uncorrelated intervals that comprise the large duration 
$\hat{\tau}$. As a result, Eq.~(\ref{relation}) can be re-cast, using 
Eqs.~(\ref{workings1}), (\ref{fluxsum}) and (\ref{replacement}), and the
derivation in Appendix \ref{canonapp}, as
\begin{equation}
\label{dyncanonical}
  \frac{\omega^{\rm dr}_{a\to b}}{\omega^{\rm eq}_{a\to b}} =
  \lim_{\tau/\tau_{\rm cor}\to\infty} \frac{\int \td J\:
  p_{\tau}^{\rm eq}(J |a\to b)\: e^{\tau\nu J} }
  {\int \td J\: p_{\tau}^{\rm eq}(J |a)\: e^{\tau\nu J} }
\end{equation}
where the control parameter $\nu$ is conjugate to the time-averaged flux, and
is fixed by the relation
\begin{equation}
\label{J}
  \frac{\d Q}{\d\nu} = J,
\end{equation}
in terms of the function
\begin{equation}
\label{Q}
  Q(\nu) \equiv \lim_{\tau\to\infty} 
  \frac{\ln \left\< e^m \right\>_{\rm dr}}{\tau}.
\end{equation}
Here, $\<\ldots\>_{\rm dr}$ is an ensemble average with respect to 
the steady-state distribution of microstates $p^{\rm dr}(c)$. We have 
defined
\begin{equation}
\label{m}
  m_c(\nu,\tau) \equiv \ln \int_{-\infty}^{\infty}
  p_{\tau}^{\rm eq}(J|c)\, e^{\tau\nu J} \td J
\end{equation}
that is a property of an instantaneous state $c$ of the system. 
Note that $m_c(\nu,\tau)$ has 
non-trivial $\tau$-dependence, containing transients for 
$\tau<\tau_{\rm cor}$, and becoming linear in $\tau$ for 
$\tau\gg\tau_{\rm cor}$, while $Q(\nu)$ is independent of $\tau$. 

The above equations have a structure that is familiar from equilibrium
thermodynamics. Clearly, in Eq.~(\ref{Q}), $Q$ plays the role of a 
thermodynamic potential, andits derivative $J$ is conjugate to the 
temperature-like parameter $\nu$.

The conditional probabilities in the integrands of
Eq.~(\ref{dyncanonical}) describe the likelihood of any particular flux
during the interval $\tau$, given the initial state and/or transition. The
exponential factor measures the change in the weight of the large
remainder of the trajectory of duration $\hat{\tau}$, due to the
initial part accepting a flux $J$ rather than postponing
it until after $\tau$.

Compare Eqs.~(\ref{dynmicro}) and (\ref{dyncanonical}). The expressions
become very similar under a change of notation
$\tau\leftrightarrow\tau_0$. The difference in the new formulation
(Eq.~(\ref{dyncanonical})) is that trajectories with the wrong flux are
not eliminated by a delta function, but re-weighted by an exponential
weight factor.

The two alternative formulations are exactly akin to alternative ensembles
in equilibrium statistical mechanics. We can regard the duration of a
trajectory as being analogous to the size of a system at equilibrium, and
the flux as analogous to energy-density. Originally we demanded that the
integrated flux was fixed exactly, just as energy is fixed in the
mico-canonical ensemble, and we enquired, in Eq.~(\ref{dynmicro}), about
how the instantaneous (`local') conditions are affected by correlations in
the rest of the trajectory (`system'). The formulation of
Eq.~(\ref{dyncanonical}) is akin to using the canonical ensemble. Again,
we enquire about conditions at an instant (`locality') $\Delta t$, within
a trajectory segment (`system') of duration (`size') $\tau$. But now, the
integrated flux (`energy') is not strictly conserved, but can be exchanged
with the rest of the trajectory (`a reservoir') of duration (`size')
$\hat{\tau}$ much longer (`larger') than the initial trajectory segment 
(`system'). Since all important correlations are contained within the
`system', the nature of the interface between `system' and `reservoir'
becomes unimportant, and the `reservoir' is characterised by a single
parameter, $\nu$. Let us refer to this as the`canonical-flux' ensemble.
So long as the trajectory duration (`system size') is much greater than
any correlation time (`length'), the properties at instant (`locality')
$\Delta t$ are unaffected by whether integrated flux (`energy') is exactly
conserved, and the ensembles are equivalent in the infinite-time
(`thermodynamic') limit.

It is possible to derive Eq.~(\ref{dyncanonical}) [via Eqs.~(\ref{eqrate})
and (\ref{drivenrate})] by direct maximization of the information entropy
of a set of trajectories, at fixed ensemble-averaged flux and energy. In
that case, as with the above derivation, great care is required to compare
the relevant time-scales with correlation times, to avoid unwittingly
averaging over the correlations present in $p_{\tau}^{\rm eq}(J|a\to b)$.
That would produce a mean-field expression, in which the rate of each
transition is simply boosted exponentially according to its immediate flux
contribution. Such a scheme is popular in simple models, but should not be
mistaken for the exact theorem derived above.

\subsection{Factors affecting transition rates}
\label{factors}

The expression for transition rates, Eq.~(\ref{dyncanonical}), appears to 
depend on the arbitrary quantity $\tau$. It can be re-written in an a much
clearer form that is explicitly independent of $\tau$, as we now show.

Although $m_c(\nu,\tau)\to\infty$ as $\tau\to\infty$, the difference
$m_b-m_a$, for two states $a$ and $b$, has a finite asymptote, embodying the 
different transient influences that 
the two states have on the system, before it returns to a statistically 
steady state. So, let us define a function that contains that transient 
information, but is independent of the arbitraryquantity $\tau$, thus:
\begin{eqnarray}
  q_a(\nu) &\equiv& \lim_{\tau\to\infty} \{ m_a(\nu,\tau)-\tau Q(\nu)\}
\label{qm}  \\
  &=& \ln\lim_{\tau\to\infty} \frac{e^{m_a(\nu,\tau)}}
  {\left\langle e^{m(\nu,\tau)}\right\rangle_{\rm dr}}  \\
  &=& \ln\lim_{\tau\to\infty} 
  \frac{\int p_{\tau}^{\rm eq}(J|a)\,e^{\tau\nu J}\,\td J}
  {\int \left\langle p_{\tau}^{\rm eq}(J)\right\rangle_{\rm dr}\,
  e^{\tau\nu J}\,\td J}
\label{willingness}
\end{eqnarray}
so that $q_a-q_b=m_a-m_b$ in the long-time limit.

We require one further piece of notation. 
The dynamics is described by a set of transitions
$a\to b$ carrying integrated flux $J_{ab}\:\Delta t$. For continuous dynamics, 
$J_{ab}\:\Delta t\to 0$ as $\Delta t\to0$, but for discrete transitions, 
$J_{ab}\:\Delta t$ remains finite whether or not time steps are made 
vanishingly small. As above, the following discussion applies to either case.

In terms of these physically meaningful quantities, transition 
rates in the driven ensemble are given by
\begin{equation}
\label{clear}
  \omega^{\rm dr}_{a\to b} = \omega^{\rm eq}_{a\to b} \: \exp\big[
  \nu J_{ab}\,\Delta t +q_b(\nu)-q_a(\nu) - Q(\nu) \Delta t\big].
\end{equation}
The derivation of Eq.~(\ref{clear}) from Eq.~(\ref{dyncanonical}) is given in 
Appendix \ref{reformapp}.
It is now clear, in Eq.~(\ref{clear}), that three distinct factors
determine the rate of a transition $a\to b$ in the driven steady-state
ensemble. (1) The rate is proportional to the rate at equilibrium. So, all
else being equal, energetically expensive transitions are slow, while
down-hill transitions take precedence. (2) The rate is exponentially
enhanced for transitions that contribute a favourable flux. (3) The
dependence on $q_b-q_a$ is overlooked by mean-field models. It says that a
transition's likelihood depends also on the state in which it leaves the
system. Its rate is enhanced if it puts the system into a state that is
more likely to exhibit flux in the future. The effect of this factor on the 
driven steady-state distribution of microstates is to increase (relative to the 
Boltzmann distribution) the weight of states that are more-than-averagely 
willing to accept a future flux. We shall see an example 
of this effect in the next section. In the case of a shear flux, this means 
that low-viscosity states are favoured, as is often observed.

\section{Applications}
\label{applications}

We have a recipe for constructing a model of any given driven system, that is 
guaranteed to yield the desired flux, and to respect all the physical laws that 
are obeyed by the equilibrium version of the model, and that is guaranteed to 
have no artefacts from statistical bias. If we choose to provide this machinery 
with an equilibrium model that obeys all of Newton's laws --- i.e., a fluid 
whose internal interactions conserve momentum, angular momentum, and energy, 
while stochastic forces from the reservoir couple only to particles at the 
boundary --- then it will yield dynamical rules that also respect Newton's laws 
for the boundary-driven fluid. In other words, the method has the capacity to 
produce exact physics if provided with an exactly physical prior. It provides a 
description of the reservoir, by characterising the stochastic part of the 
equations of motion. The way in which that reservoir couples to the system is up 
to the user to decide. In the above example, it is coupled only at the boundary, 
but we may instead consider a driven Brownian system, for which the heat bath is 
more strongly and uniformly coupled, dominating all momentum variables. Another 
alternative is to apply the method to a model whose prior (equilibrium) physical 
properties are simplified for the sake of clarity and analytical expediency. In 
that case, of course, the result of the recipe will be approximate and 
unreliable, but still the least arbitrary choice of transition rates for the 
given degree of simplification.

The micro-canonical flux ensemble introduced in sections \ref{presentation} and 
\ref{micro} was first presented in Ref.~\cite{PRL}, where it was used 
analytically to construct a continuum model of driven diffusion, and 
heuristically to discuss the features of a lattice model of dimers under shear. 
The latter model had a much more complex energy landscape including jammed 
states. Another analytically solvable model was constructed in 
Ref.~\cite{sardinia}, using the micro-canonical flux method. It was another 
one-dimensional driven diffusion model, but this time with a discrete state 
space and discrete time step. In the following section, we shall analytically 
construct a model of a driven system with a non-trivial energy landscape, that 
demonstrates some features of more complex systems, such as sheared complex 
fluids, with states that are locally trapped so that they cannot easily be 
driven. The model reduces to simple one-dimensional driven diffusion in a 
certain limit, and has a discrete state space but continuous time, to complement 
the earlier published models. We shall use the canonical-flux ensemble of 
sections \ref{canon} and \ref{factors}, to demonstrate the utility of this 
method.

\subsection{The model}
\label{model}

\begin{figure}
  \begin{center}
  \epsfxsize=7.5cm
  \leavevmode\epsffile{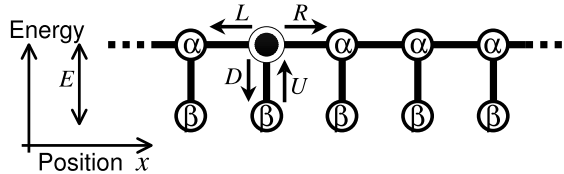}
  \caption{Comb-shaped state-space for the hopping model. Circles represent the 
possible states of the system, of types $\alpha$ and $\beta$. The single 
particle, shown as a filled circle, currently occupies an $\alpha$ state. It may 
escape into other $\alpha$ states to the left or right with rates $L$ and $R$ 
respectively, or downwards in energy, with rate $D$, to state $\beta$. Each 
$\beta$ state can be exitted with rate $U$ upwards to the $\alpha$ state at the 
same location only.}
  \label{comb}
  \end{center}
\end{figure}

Consider a particle that can hop stochastically among a set of discrete states 
that have the connectivity shown in Fig.~\ref{comb}. The particle will be driven 
by its non-equilibrium heat bath so that it has, on average, a drift velocity 
$u$ from left to right. At each location $x$, it may occupy one of two states: 
state $\alpha$, from which it may escape to the left or right with rates $L$ and 
$R$ to other $\alpha$ states at different locations $x$, or downwards with rate 
$D$ into a lower energy trapped state $\beta$; once in a $\beta$ state, the 
particle cannot exhibit any flux, i.e., cannot move left or right, but can only 
wait for a random excitation at rate $U$ back up to the $\alpha$ state at the 
same location.

Note that, if we set $D=0$, the model reduces to a continuous-time 
discrete-space linear hopping model, like the versions that were 
previously studied with 
both space and time continuous \cite{PRL} or discrete \cite{sardinia}.

The equilibrium version of this model (with no mean drift) is very 
straightforward. Detailed balance requires that 
$U^{\rm eq}=D^{\rm eq}\,\exp(-E)$, where E is the energy difference between 
states $\alpha$ and $\beta$ 
measured in units of $k_BT$, and that $R^{\rm eq}=L^{\rm eq}\equiv\omega_0$ 
where we may measure all rates in units of $\omega_0$ so that $\omega_0\equiv1$ 
without loss of generality. The occupancy of $\alpha$ states is given by 
Boltzmann as $1/(1+\exp(E))$, and the only remaining parameter that we are free 
to choose is $D^{\rm eq}\equiv\rho$, which specifies the ratio of vertical to 
horizontal mobilities.

When the model is not at equilibrium, but is driven at drift velocity $u$, the 
na\"{\i}ve expectation would be either that we a free to choose all four rates 
$U, D, L, R$, since non-equilibrium models traditionally have no rules, or that 
detailed balance still governs the ratio $U/D$. However, as discussed above, 
neither of these statements is true. There is, in fact, a least-arbitrary set of 
rates, that corresponds to driving by an uncorrelated heat bath that is 
characterised only by its temperature and velocity. We shall now calculate that 
set of rates, using the canonical-flux ensemble.

The rates are given by Eq.~(\ref{clear}). Defining our unit of length to be one 
inter-site spacing, the integrated flux of a transition to the right (left) is 
$J_R\,\Delta t=1$, ($J_L\,\Delta t=-1$), while transitions between $\alpha$ and 
$\beta$ states carry no flux as they leave the particle's displacement 
unchanged. Since time is continuous, the time-step is infinitesimal, $\Delta 
t\to 0$, so that the last term in the exponential of Eq.~(\ref{clear}) vanishes, 
and it prescribes the following rates in the driven ensemble:
\begin{subequations}
\label{RLDU}
\begin{eqnarray}
  R &=& e^\nu
\label{R}  \\
  L &=& e^{-\nu}
\label{L} \\
  D &=& \rho\,e^{q_{\beta}-q_{\alpha}}
\label{D}  \\
  U &=& U^{\rm eq}\,e^{q_{\alpha}-q_{\beta}}
  = \rho\,e^{-E+ q_{\alpha}-q_{\beta}}.
\label{U}
\end{eqnarray}
\end{subequations}
To complete the calculation of the rates, we require only 
$q_{\alpha}-q_{\beta}$, the difference in the willingness of the two 
states to admit a flux. 
This could be evaluted by ``brute force" using Eq.~(\ref{willingness}), if we 
first calculate the Green function for the equilibrium model, i.e.~the 
probability that the particle travels a given distance in a given time, given 
the intial state $\alpha$ or $\beta$. However, that calculation can be avoided, 
using the derivation in Appendix \ref{combapp} to show that, for this ``comb" 
model,
\begin{equation}
\label{qq}
  q_{\alpha}(\nu)-q_{\beta}(\nu) = \ln(1+Q(\nu)/U^{\rm eq}).
\end{equation}
This is purely a result of the facts that state $\beta$ can only be quitted via 
state $\alpha$, and that escape times are distributed exponentially. 

We can now construct a differential equation for $Q(\nu)$, as follows. Due to 
the model's translational symmetry, the steady-state occupancy of $\alpha$ 
states is just
\begin{equation}
\label{occ}
  f_{\alpha} = \frac{U}{U+D}
\end{equation}
and, since displacements are allowed only from $\alpha$ states, the mean drift 
velocity is
\begin{equation}
\label{drift}
  u = (R-L)\,f_{\alpha} = 2f_{\alpha}\,\sinh\nu.
\end{equation}
Now, using Eq.~(\ref{J}), we obtain an ordinary differential equation,
\[
  \frac{\td Q}{\td\nu} = \frac{2(\rho\,e^{-E}+Q)^2\sinh\nu}
  {\rho^2 e^{-E}+(\rho\,e^{-E}+Q)^2}
\]
that can be integrated for $\cosh\nu$ as a function of $Q$.
The constant of integration is fixed by $Q(0)=0$ which follows from 
normalization of the probability distribution in the definition of $Q$ 
[Eqs.~(\ref{Q}) and (\ref{m})]. Finally, we obtain the required ``potential",
\[
  Q(\nu) = \cosh(\nu)-1 - \frac{\rho}{2}(1+e^{-E})
  + \sqrt{\left[ \cosh(\nu)-1-\frac{\rho}{2}(1-e^{-E})
  \right]^2 +\rho^2 e^{-E}}
\]
which, with Eqs.~(\ref{RLDU}), (\ref{qq}), (\ref{occ}) and (\ref{drift}), leads 
to four constraints on the four transition rates in the driven system, from 
which the abstract quantities $\nu$ and $Q$ have been eliminated:
\begin{subequations}
\label{constraints}
\begin{eqnarray}  
  R-L &=& (1+ D/U) \,u
\label{c1}  \\
  R\,L &=& 1
\label{c2}  \\
  U\,D &=& \rho^2 e^{-E}
\label{c3}  \\
  R+L+D-U &=& 2+(1-e^{-E}) \rho
\label{c4}
\end{eqnarray}
\end{subequations}
One of these four equations is obvious; the others are not. Equation (\ref{c1}) 
is simply a re-statement of Eq.~(\ref{drift}), and gives the drift velocity $u$ 
that results from any choice of the four transition rates. So, if we applied the 
usual {\em ad hoc} construction of non-equilibrium stochastic models, we would 
pluck four rates out of the air, use Eq.~(\ref{c1}) to find the resulting drift 
velocity, and have no other constraints. The other three constraints have arisen 
from our demand that the design of the model incorporates the prior dynamics, 
the large-scale flux, and no other design features.

Note that Eqs.~(\ref{c2}) and (\ref{c3}) express relations between forward and 
reverse transition rates that are generic to any continuous-time model with 
instantaneous transitions, that {\em the product of the forward and reverse 
rates of a transition is equal in the driven and equilibrium ensembles}. This 
follows directly from Eq.~(\ref{clear}) with $J_{ab}\,\Delta t$ finite as 
$\Delta t\to 0$.

The four transition rates defined by Eqs.~(\ref{constraints}) have exactly the 
same number of free parameters as in an equilibrium model: for a given energy 
gap $E$ and flux $u$, the four rates are defined up to one parameter, $\rho$, 
that specifies the prior ratio of vertical to horizontal mobilities, as is the 
case in the equilibrium version of the model that was required to respect 
detailed balance. While the equilibrium occupancy, given by Boltzmann's law, is 
independent of $\rho$, the occupancy in the driven ensemble [(Eq.~(\ref{occ})] 
does depend on this kinetic parameter.

\subsection{Properties of the model}

\begin{figure}
  \begin{center}
  \epsfxsize=7.5cm
  \leavevmode\epsffile{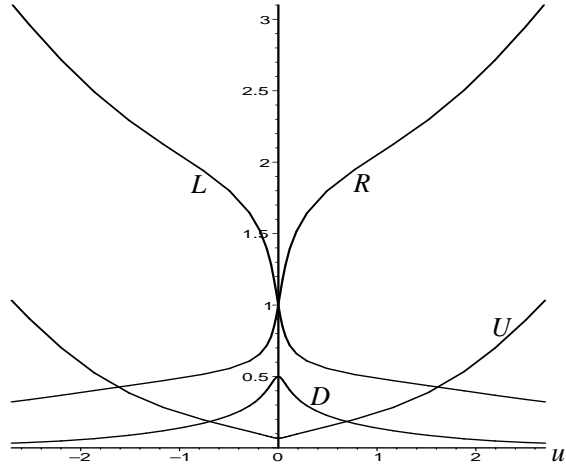}
  \caption{Rates $R$, $L$, $U$ and $D$ for the driven comb model, as functions 
  of the drift velocity $u$, using the parameters $E=2$, $\rho=0.5$.}
  \label{rates}
  \end{center}
\end{figure}

The transition rates prescribed by sub-ensemble dynamics are plotted as 
functions of velocity $u$ in Fig.~\ref{rates} for an energy gap $E=2$ and 
mobility ratio $\rho=0.5$. Due to the symmetries of the comb structure, the 
rates of transitions up and down ($U$, $D$) between $\alpha$ and $\beta$ states 
are even functions of $u$. At $u=0$, the rates take their equilibrium values, 
$R=L\equiv1$ and $D=U\exp E$. On increasing velocity, hops to the right ($R$) 
become more frequent, while hops to the left ($L$) are suppressed, as expected. 
Also the particle becomes less likely to fall down ($D$) into a trapped $\beta$ 
state, and is increasingly dragged out of traps ($U$) by the driving force.

\begin{figure}
  \begin{center}
  \epsfxsize=7.5cm
  \leavevmode\epsffile{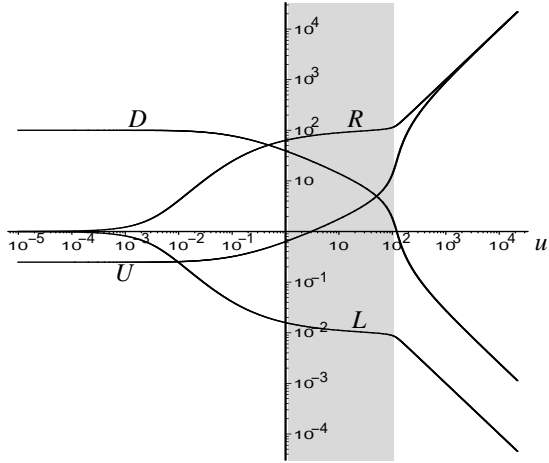}
  \caption{The transition rates of the driven comb model, shown on log-log axes 
  as functions of the positive drift velocity $u$. Parameter values are $E=6$, 
  $\rho=100$. Three regimes of behaviour are visible. The regime at
  intermediate values of $u$ is shaded.}
  \label{logrates}
  \end{center}
\end{figure}

The rates are re-plotted on log-log axes (for positive $u$) in 
Fig.~\ref{logrates}, using parameter values $E=6$, $\rho=100$, that were chosen 
to provide a separation of time-scales, emphasizing the features of the graphs. 
Three regimes of drift velocity $u$ become apparent. On the left of the figure 
(low $u$) is the near-equilibrium regime, where the rates $D$, $U$, of 
transitions that do not carry a flux, remain approximately constant, respecting 
detailed balance. This fulfils the na\"{\i}ve expectation often applied to 
non-equilibrium models, that detailed balance continues to describe the 
physics of 
activated processes. Meanwhile, the rate of hops to the right, $R$, is enhanced 
and to the left, $L$, is suppressed, so that the sparsely populated $\alpha$ 
states exhibit the required drift velocity.

\begin{figure}
  \begin{center}
  \epsfxsize=7.5cm
  \leavevmode\epsffile{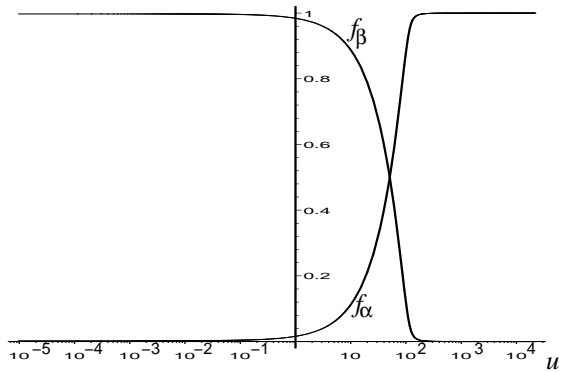}
  \caption{Occupancies of $\alpha$ and $\beta$ states for the same parameters 
  as in Fig.~\protect\ref{logrates}, shown as functions of the drift velocity, 
on 
  semi-logarithmic axes. Three zones of behaviour are again visible:
  Boltzmann-distributed states at low $u$; re-population of states at 
  intermediate $u$; the mobile $\alpha$ state fully occupied at high $u$.}
  \label{fafb}
  \end{center}
\end{figure}

The second regime of the driving velocity, $\omega_0<u<\rho$ is shaded grey in 
Fig.~\ref{logrates}. In this regime, the flux constraint can no longer be 
satisfied by the small population of thermally-activated $\alpha$ states. The 
states become mechanically activated, with particles in the immobile $\beta$ 
state promoted into the mobile state by the driving force. As $u$ increases 
through the shaded part of the figure, the unequal hopping rates to the right 
and left remain approximately constant, while the rate of activation $U$ 
increases and rate of trapping $D$ decreases, so that the drifting $\alpha$ 
population increases. This is also apparent in Fig.~\ref{fafb}, which shows the 
occupancies of the two states as a function of velocity, for this same set of 
parameters.

Once the mobile states are fully populated, and the trapped states have 
negligible occupancy, the bias on hops to the right can no longer remain 
constant while satisfying an increasing flux constraint. Hence, a third regime 
exists at the highest values of $u$ (Fig.~\ref{logrates}), where rates $R$ and 
$U$ both become proportional to the flux $u$, while the flux-impeding 
transitions have rates $L$ and $D$ inversely proportional to $u$.

\subsection{Comparison with shear flow}

The model we have studied here, with its simple comb-shaped state space, has 
some features that are generic to driven systems. We saw, in section 
\ref{factors}, that the rate of a transition in a driven ensemble depends on 
three factors: its rate at equilibrium, the amount of flux that it 
contributes, and the difference in the willingness of the initial and final 
states to allow the required flux in the future. Transitions that contribute
a non-zero amount of flux were called ``type A" in Ref.~\cite{PRL}, while 
transitions between states with different promise for future flux where labelled
``type B". In previous articles \cite{PRL,sardinia}, the rates prescribed by 
sub-ensemble dynamics were calculated explicitly only for simple models, that
exhibited only type A transitions due to the simplicity of their state spaces. 
The comb model, on the other hand, has both type A ($\alpha\to\alpha$) and 
type B ($\alpha\rightleftharpoons\beta$) transitions. Another example of such 
a model, that was previously discussed only heuristically \cite{PRL,sardinia}, 
is a set of dimers (particles that occupy two adjacent lattice sites) that 
perform random walks on a two-dimensional triangular lattice, while the lattice 
itself is 
driven into shear flow by sporadically cleaving and re-positioning its 
horizontal layers. Certain arrangements of the dimers (analogous to $\alpha$ 
states of the comb model) allow these quanta of shear, while other states
(analogous to $\beta$ states) are prevented from shearing, due to the 
unbreakable dimers straddling two layers of the lattice, thus jamming the 
system. 

Although any such many-particle system has a very complex
state-space, its crucial features are reproduced in the comb model. 
When the comb model is stuck in a $\beta$ state, the driving force (that 
derives from the statistics of the driven ensemble) pushes it 
into a more mobile state in order to flow. Likewise, when the dimer model is 
in a state that will not admit a flux, it must first re-arrange its particles. 
The driving force achieves this by imposing a shear stress on the particles, 
causing them to re-orient mechanically (as opposed to thermally, by Brownian 
motion). The sub-ensemble rules prescribe (for a given prior dynamics) the 
rate of that mechanically imposed re-alignment, thereby specifying the 
constraints that must be met by a physically acceptable constitutive relation 
for the flowing system.

\section{Summary and outlook}

There are certain constraints that must be satisfied by any candidate for a 
statistical mechanical theory of driven steady states: it must satisfy the
known laws of motion, and it must give rise to the required macroscopic 
observables (flux, energy etc.). In this article, we have assumed; indeed, 
demanded; that those are the {\em only} constraints, and derived the 
transition rates implied by that assumption. Comparison with experimental 
observations will determine {\em a posteriori} whether a particular system 
belongs to the ergodic class that is well described by these unbiased rates,
just as empirical comparison determines whether or not a static system is at
thermodynamic equilibrium. If one is privy to prior information indicating 
that the driven system's motion is biased in some way that is not apparent 
in its macroscopic flux and conserved quantities, then the dynamical rules set 
out here should be disregarded. To violate the rules {\em a priori} without 
such a justification is to bias the model with arbitrary information derived 
from prejudice rather than from physics. Such arbitrariness is not condoned 
for equilibrium models, and the same should be the case for macroscopically 
driven steady states.

For example, consider how we design a stochastic model of an equilibrium 
system. The system is defined by some set of available states, and we must
choose the rates of transitions between those states. Canonical equilibrium is
defined by a fixed volume, particle number, and mean energy of the system. We 
might choose any arbitrary set of rates, and then measure or calculate the mean
energy that results when the system arrives at a steady state. Certainly, that
procedure would give rise to a well defined mean energy, volume and 
number, but that is not sufficient for us to say that the system is at 
equilibrium and that the transition rates are acceptable. There are 
constraints arising from the principle of detailed balance, which ensure that
$(E, N, V)$ are the {\em only} parameters characterising the macroscopic state
of the ensemble, beyond the definition of the system in terms of its accessible
states and reversibility. We have found the generalisation of those constraints 
to non-equilibrium steady states.

The prior is central to the formalism, and is often misinterpreted in
non-equilibrium applications of information theory. In the present context, it
is used to mean the complete set of {\em physically valid} paths that a system
might take in response to the stochastic forces arising from a particular 
coupling to a non-equilibrium reservoir. If the reservoir can exchange energy
with the system, then conservation of energy can be violated in the prior. If
the coupling is only to particles at the system's boundary, then energy, 
momentum and angular momentum must be conserved by all internal interactions,
so the prior does not include scenarios for which those laws are violated.

This has not been the usual definition of the prior, in previous attempts at
non-equilibrium applications of information theory. It is often assumed that
our knowledge of microscopic dynamics can be discarded, and MaxEnt will 
correctly reconstruct that missing information. Such optimism cannot be 
justified. For instance, MaxEnt has been used to choose between phase-space 
paths that are characterised by their actions \cite{Wang}, discarding our 
knowledge of Hamilton's principle of least action. The result is an exponential 
distribution in which the paths of least action are the most likely, but that 
result is incorrect. Paths on which the action is extremized are not just {\em 
likely}; they are the {\em only} paths of a classical system, and therefore the 
only paths that should appear in the prior if an exact calculation is wanted.

The central results of this paper are the formulae for transition rates in
a driven ensemble. These are formulated in two alternative ways. In the
``microcanonical-flux ensemble", the flux is constrained to an exact value
when time-averaged over the duration (tending to infinity) of each
system's passage through phase-space, resulting in Eq.~(\ref{relation}).
The ``canonical-flux ensemble", in which only the ensemble-averaged flux is 
constrained, leads to Eq.~(\ref{clear}) for the transition rates, which
is exactly equivalent to the microcanonical-flux prescription. The
canonical-flux equation (\ref{clear}) 
makes explicit the three factors influencing a transition rate. As at
equilibrium, energetics are important, making a system reluctant to take
up-hill steps in its energy landscape. Secondly, an exponential factor, that
one might have guessed, favours transitions that impart the desired
flux. The third factor prescribed by Eq.~(\ref{clear}) is more subtle. 
It describes the importance of correlations, and depends on a well-defined 
quantity ascribed to each microstate, that quantifies its promise for future 
flux. A transition is favoured if it takes the system to a state of higher 
promise, that is more amenable to future flux-carrying transitions.

The sub-ensemble scheme has previously been demonstrated to produce the 
standard equations
of motion for diffusion with drift, both for continuous \cite{PRL} and
discrete \cite{sardinia} random walks. In the present article, the dynamical
rules were evaluated for a more complex model. We have seen that, for thermally
activated processes, that are governed by detailed balance at equilibrium, the
sub-ensemble rules describe {\em mechanical} activation by the driving force,
although detailed balance is recovered in the low flux regime. 

In the context of shear flow, mechanical activation  corresponds to 
stress-induced re-arrangement. The fact that this statistical formalism 
describes the effects of non-equilibrium stresses in a natural way, makes it a
promising approach for the study of shear-banding, jamming, and other 
shear-induced transitions of complex fluids.

At the risk of repetition, we have a recipe for constructing a model of any 
given driven system, that is guaranteed to yield the desired flux, and to 
respect all the physical laws that are obeyed by the equilibrium version of the 
model. It is also guaranteed to have no artefacts from statistical bias. This 
machinery has the capacity to produce exact physics if provided with an exactly 
physical prior. Otherwise, it will yield the least arbitrary model for the given 
degree of approximation.

\section{Acknowledgments}
Many thanks go to Alistair Bruce, Michael Cates, Richard Blythe, Peter
Olmsted, Tom McLeish, Alexei Likhtman, Suzanne Fielding and Hal Tasaki for
informative discussions. RMLE is grateful to the Royal Society for
support.

\appendix

\section{The canonical-flux potentials}
\label{canonapp}

As stated in Eq.~(\ref{replacement}), the distribution of flux $\hat{J}$ 
during interval $\hat{\tau}$, that appears in Eq.~(\ref{workings1}), is 
uncorrelated with the initial state $a$, and can therefore be written
\begin{equation}
\label{f}
  p_{\hat{\tau}}(\hat{J}|a, J) \to f_{\hat{\tau}}(\hat{J}) \equiv 
  \sum_c p^{\rm dr}(c) \:p_{\hat{\tau}}^{\rm eq}(\hat{J}|c)
\end{equation}
in terms of the instantaneous steady-state distribution of states
$p^{\rm dr}(c)$. The distribution
$f_{\hat{\tau}}(\hat{J})$ can be evaluate if we sub-divide the interval
$\hat{\tau}$ into $n$ sub-intervals of duration $\tau$, where $n\gg1$
since $\hat{\tau}\sim\tau_0\gg\tau$. The system begins each of these
sub-intervals in a state drawn randomly and independently from the
steady-state distribution $p^{\rm dr}(c)$. These initial states are
uncorrelated because $\tau\gg\tau_{\rm cor}$. The overall flux in the
interval $\hat{\tau}$ is the mean of the fluxes in these $n$ independent
sub-intervals, so that
\begin{equation}
\label{convolution}
  f_{\hat{\tau}}(\hat{J}) = \int_{-\infty}^{\infty} \td J_1\ldots\td J_n\;
  f_{\tau}(J_1)\ldots f_{\tau}(J_n)\;
  \delta(\hat{J}-\frac{1}{n}\sum_i^n J_i).
\end{equation}
This limit distribution gives the likelihood (under equilibrium dynamics,
with a non-equilibrium initial state) that the $n$ independent flux
measurements have an improbably-large mean value $\hat{J}$. Cam\'{e}r's
theorem of large deviations \cite{Varadhan} states that the weight in the
tail of the distribution of the mean of $n$ independent identically
distributed random variables behaves as
\begin{equation}
\label{cramer}
  \lim_{n\to\infty} \frac{1}{n}\ln\int_{\hat{J}}^{\infty} f_{n\tau}(J')\,
  \td J' = -I(\hat{J},\tau).
\end{equation}
That is, the weight in the tail decays exponentially with $n$, at a rate
$I$ given \cite{Varadhan} by
\begin{equation}
\label{sup}
  I(J,\tau) = \sup_{\theta}\left[ \theta J - 
  \ln\int_{-\infty}^{\infty} f_{\tau}(J')\, e^{\theta J'}\,\td J' \right].
\end{equation}
Dividing both sides of Eq.~(\ref{cramer}) by the constant $\tau$ gives
\begin{equation}
\label{limit}
  \lim_{\hat{\tau}/\tau\to\infty} \frac{1}{\hat{\tau}}
  \ln\int_{\hat{J}}^{\infty} f_{\hat{\tau}}(J')\,\td J'
  = \frac{-I(\hat{J},\tau)}{\tau}.
\end{equation}
Since the LHS of Eq.~(\ref{limit}) is independent of the arbitrary choice
of $\tau$, we can infer that $I\propto\tau$. Let us define the function
\begin{equation}
\label{defH}
  H(J) \equiv I(J,\tau)/\tau
\end{equation}
that is independent of the arbitrary quantity $\tau$. Writing the 
exponential decay law explicitly, with an unknown prefactor $A(\hat{J})$ 
that varies only slowly with $\hat{\tau}$,
\[
  \int_{\hat{J}}^{\infty} f_{\hat{\tau}}(J')\, \td J'
  \to A(\hat{J})\,\exp\left[ -\hat{\tau} H(\hat{J}) \right]
  \;\mbox{ as }\; \tau/\hat{\tau}\to 0 
\]
and differentiating with respect to $\hat{J}$ gives
\begin{equation}
\label{exp}
  f_{\hat{\tau}}(\hat{J})\to \left[ 
  A(\hat{J})\hat{\tau} H'(\hat{J}) - A'(\hat{J}) \right]
  \exp\left[ -\hat{\tau} H(\hat{J}) \right].
\end{equation}
Now, substituting for $\hat{J}$ from Eq.~(\ref{fluxsum}) and
Taylor-expanding $H(\hat{J})$ to first order in $\tau/\hat{\tau}$ allows
us to take the limit $\hat{\tau}/\tau\to\infty$ when substituting
Eqs.~(\ref{workings1}), (\ref{f}) and (\ref{exp}) into (\ref{relation}),
yielding
\begin{equation}
\label{workings2}
  \frac{\omega^{\rm dr}_{a\to b}}{\omega^{\rm eq}_{a\to b}} =
  \lim_{\tau/\tau_{\rm cor}\to\infty} \frac{\int_{-\infty}^{\infty}\td J\:
  p_{\tau}^{\rm eq}(J |a\to b)\: e^{\tau\,H'(J_0)\,J} }
  {\int_{-\infty}^{\infty} \td J\: p_{\tau}^{\rm eq}(J |a)\: 
  e^{\tau\,H'(J_0)\,J} }.
\end{equation}

The supremum in Eq.~(\ref{sup}) can be evaluated by defining the functions
in Eqs.~(\ref{Q}) and (\ref{m}). From 
Eqs.~(\ref{f}), (\ref{sup}) and (\ref{defH}), we have
\begin{equation}
\label{H}
  H(J) = \nu J - Q(\nu)
\end{equation}
with the parameter $\nu(J)$ [equal to $\theta/\tau$ in Eq.~(\ref{sup})]
given by Eq.~(\ref{J}). Thus the parameter $H'(J_0)$ in Eq.~(\ref{workings2}) 
can be evaluated by differentiating Eq.~(\ref{H}) and substituting from 
Eq.~(\ref{J}), to give
\begin{equation}
\label{theta}
  \frac{\d H}{\d J} = \nu
\end{equation}
resulting in Eq.~(\ref{dyncanonical}). Note that $Q$ is a Legendre transform of
$H$, and that $\nu$ and $J$ in Eqs.~(\ref{theta}) and (\ref{J}) are conjugate
variables.

\section{Re-formulation of the canonical flux expression for transition rates}
\label{reformapp}

Let us make a change of variable in Eq.~(\ref{dyncanonical}), and replace
the integration over {\em average} flux $J$ by one over {\em total}
(integrated) flux $K\equiv\tau J$. Then, using 
$p^{\rm eq}_{\tau}(K|\ldots)\,\td K$ now to represent the normalized 
probability of finding a total flux $K$ on an equilibrium trajectory of 
length $\tau$, we can write
\begin{equation}
\label{rewrite}
  \frac{\omega^{\rm dr}_{a\to b}}{\omega^{\rm eq}_{a\to b}} =
  \lim_{\tau/\tau_{\rm cor}\to\infty} \frac{\int \td K\:
  p_{\tau}^{\rm eq}(K |a\to b)\: e^{\nu K} }
  {\int \td K\: p_{\tau}^{\rm eq}(K |a)\: e^{\nu K} }.
\end{equation}
Now, the expression $p_{\tau}^{\rm eq}(K|a\to b)$ is the probability of 
accumulating an integrated flux $K$ during interval $\tau$, given that 
the initial part $\Delta t$ of that interval is taken up with a 
transition from state $a$ to $b$. Since that transition carries an 
integrated flux $K_{ab}\equiv J_{ab}\:\Delta t$, we can replace the 
expression by the probability of accumulating the remaining flux $K-K_{ab}$ 
in the remaining time, starting from state $b$, i.e.
\[
  p_{\tau}^{\rm eq}(K|a\to b) = p_{\tau-\Delta t}^{\rm eq}
  \left(\left. K - K_{ab} \; \right| \;b \right).
\]
Hence, after a change of variable, Eq.~(\ref{rewrite}) gives
\begin{eqnarray}
\label{line1}
  \ln\frac{\omega^{\rm dr}_{a\to b}}{\omega^{\rm eq}_{a\to b}} 
  &=& \nu K_{ab} + \lim_{\tau\to\infty}
  \left[ m_b(\nu,\tau-\Delta t)-m_a(\nu,\tau) \right]  \\
\label{line2}
  &=& \nu K_{ab} + \lim_{\tau\to\infty} \left[ 
  m_b(\nu,\tau)-m_a(\nu,\tau) \right] - \zeta_b(\nu,\Delta t)
\end{eqnarray}
where
\begin{equation}
\label{zeta}
  \zeta_b(\nu,\Delta t) \equiv \lim_{\tau\to\infty}
  \left[ m_b(\nu,\tau)-m_b(\nu,\tau-\Delta t) \right]
\end{equation}
and, by substituting $\tau\to\tau+\Delta t$ into Eq.~(\ref{line1}), we
find $\zeta_a=\zeta_b\;\forall\;a,b$, i.e., $\zeta_b$ is 
state-independent. Given that the limit in Eq.~(\ref{zeta}) exists, we
can write
\begin{equation}
  \zeta_b(\nu,\Delta t) = \Delta t \lim_{\tau\to\infty}
  \left(\frac{\d m_b}{\d\tau} \right)_{\nu}
\end{equation}
even for finite $\Delta t$, since $m_b$ asymptotes to a linear function 
of $\tau$. The state-independence of $\zeta_b$ can now be used to factor 
out the time-derivate of $m$ from the ensemble average when 
differentiating Eq.~(\ref{Q}) with respect to $\tau$, yielding
\begin{equation}
\label{dmadt}
  \zeta_b(\nu,\Delta t) = Q(\nu)\,\Delta t \; \forall \; b.
\end{equation}
Finally, substituting Eq.~(\ref{dmadt}) into (\ref{line2}) gives a 
very simple expression for the ratio of transition rates,
\begin{equation}
\label{nearlythere}
  \ln\frac{\omega^{\rm dr}_{a\to b}}{\omega^{\rm eq}_{a\to b}} =
  \nu K_{ab} - Q(\nu) \Delta t +
  \lim_{\tau\to\infty} \left[ m_b(\nu,\tau) - 
  m_a(\nu,\tau) \right]
\end{equation}
from which Eq.~(\ref{clear}) follows.

\section{Calculation for continuous-time hopping on a comb}
\label{combapp}

For the discrete states of the comb model of section \ref{model}, with the 
integrated flux $J\,\Delta t$ quantized into discrete values of the displacement 
$x$, Eq.~(\ref{m}) becomes
\[
  m_{\beta}(\nu,\tau) = \ln \sum_{x=-\infty}^{\infty} G_{\beta}(x,\tau)\,
  e^{\nu x}
\]
where $G_{\beta}(x,\tau)$ is the equilibrium Green function for $\beta$ states. 
That is the probability of attaining a displacement $x$ in time $\tau$ given 
that the particle initially occupies a $\beta$ state. An equivalent expression 
holds for $m_{\alpha}$.

In the continuous-time model, a particle occupying state $\beta$ at time $0$ 
will escape to the corresponding $\alpha$ state at a time $t$ that is drawn 
stochastically from the exponential probability distribution 
$p_{\beta\alpha}(t)=U^{\rm eq}\,\exp(-U^{\rm eq}t)$. Once excited to the 
$\alpha$ state, the particle is governed by the corresponding Green function 
$G_{\alpha}(x,\tau)$, so that the Green function for a particle occupying state 
$\beta$ is given by
\begin{eqnarray*}
  G_{\beta}(x,\tau) &=& \int_0^\tau p_{\beta\alpha}(\tau-t)\: G_{\alpha}(x,t)\:
  \td t  \\
  &=& U^{\rm eq} \, e^{-U^{\rm eq}\tau} \int_0^\tau e^{U^{\rm eq}t}\,
 G_{\alpha}(x,t)\: \td t
\end{eqnarray*}
from which it follows that
\[
  e^{U^{\rm eq}\tau+m_{\beta}(\nu,\tau)} = U^{\rm eq} \int_0^\tau
  e^{U^{\rm eq}t+m_{\alpha}(\nu,t)} \td t.
\]
Differentiating with respect to $\tau$ yields
\[
  U^{\rm eq} + \frac{\d m_{\beta}(\nu,\tau)}{\d\tau} =
  U^{\rm eq} e^{m_{\alpha}(\nu,\tau)-m_{\beta}(\nu,\tau)}.
\]
In the limit of large $\tau$, the time derivative of $m_{\beta}$ is just $Q$, as 
given by Eq.~(\ref{dmadt}), so that, with the definition of $q_{\alpha}$ in 
Eq.~(\ref{qm}), the required result, Eq.~(\ref{qq}) follows.

\end{document}